\documentclass[compsoc,conference,a4paper,10pt,times]{IEEEtran}

\usepackage{booktabs}
\usepackage[colorlinks,urlcolor=blue,linkcolor=blue,citecolor=blue]{hyperref}
\usepackage{mathtools}
\usepackage{comment}
\usepackage{url}
\usepackage{tikz}
\usepackage{pgfplots}
\usepackage{setspace}
\usepackage{mathrsfs}
\usepackage{multirow}
\usepackage{multicol}
\usepackage{color,array}
\usepackage{amsfonts,amssymb}
\usepackage{makecell}
\usepackage[ruled,linesnumbered]{algorithm2e}
\usepackage{graphicx}
\usepackage{tikz}
\usepackage{subcaption}
\usepackage{pgfplots}
\usepgfplotslibrary{groupplots}
\newtheorem{definition}{Definition}[section]

\makeatletter
\newcommand{\linebreakand}{
  \end{@IEEEauthorhalign}
  \hfill\mbox{}\par
  \mbox{}\hfill\begin{@IEEEauthorhalign}
}
\makeatother

\begin{document}

\title{{Hide me Behind the Noise:} \\ Local Differential Privacy for Indoor Location Privacy}

\author{
\IEEEauthorblockN{Hojjat Navidan\IEEEauthorrefmark{1}, Vahideh Moghtadaiee\IEEEauthorrefmark{2}, Niki Nazaran\IEEEauthorrefmark{2}, Mina Alishahi\IEEEauthorrefmark{3}}
\IEEEauthorblockA{{\IEEEauthorrefmark{1}School of Electrical and Computer Engineering, University of Tehran, Tehran, Iran} \\
{\IEEEauthorrefmark{2}Cyberspace Research Institute, Shahid Beheshti University, Tehran, Iran }\\
{\IEEEauthorrefmark{3}Department of Computer Science, Open Universiteit, The Netherlands } \\
h.navidan@ut.ac.ir, v$\_$moghtadaiee@sbu.ac.ir, mina.sheikhalishahi@ou.nl }}

\maketitle        

\begin{abstract}
The advent of numerous indoor location-based services (LBSs) and the widespread use of many types of mobile devices in indoor environments have resulted in generating a massive amount of people's location data. While geo-spatial data contains sensitive information about personal activities, collecting it in its raw form may lead to the leak of personal information relating to the people, violating their privacy. This paper proposes a novel privacy-aware framework for aggregating the indoor location data employing the Local Differential Privacy (LDP) technique, in which the user location data is changed locally in the user's device and is sent to the aggregator afterward. Therefore, the users' locations are kept hidden from a server or any attackers. The practical feasibility of applying the proposed framework is verified by two real-world datasets. The impact of dataset properties, the privacy mechanisms, and the privacy level on our framework are also investigated. The experimental results indicate that the presented framework can protect the location information of users, and the accuracy of the population frequency of different zones in the indoor area is close to that of the original population frequency with no knowledge about the location of people indoors.

\end{abstract}

\section{Introduction} \label{sec:intro}
Nowadays, a large number of Internet-connected mobile devices, such as smartphones, tablets, and smart wearables, frequently generate location data. Sharing and utilizing this location data, particularly in indoor environments, is beneficial for both users and Location Service Providers (LSPs)~\cite{8493532}\cite{8486221}. To name a few, the indoor location data can be used \emph{(i)} to increase buildings safety, \emph{i.e.}, by determining the number of individuals in each area, the position of emergency exits, fire extinguishers, security alarms, and first aid boxes can be estimated; \emph{(ii)} to improve marketing revenues, \emph{i.e.}, by recognizing users' interests in shopping centers or malls and monitoring their stays in specific areas; and  \emph{(iii)} to better locate the Location-Based Services (LBSs) in crowded places such as airports,  train stations,  exhibitions, and health centers~\cite{zafari}. 

Given the utility of indoor location data, several indoor localization methods have been proposed in the literature. The most commonly used approach is the fingerprinting technique, in which a specific environmental characteristic is used to determine the location of the target.
Particularly, in the case of Wi-Fi fingerprinting, the Received Signal Strength Indicator (RSSI) measurements from Wi-Fi routers are utilized \cite{perez-navarroChallengesFingerprintingIndoor}, which can accurately determine the \emph{exact} location of users in an indoor environment. 

Determining and monitoring the users' precise location, however, threatens the users' privacy. This data might reveal users' social habits, traditions, religious attitudes, workplaces, travel times, and health conditions \cite{Alhalafi2019}. This confidential information, for instance, can be exposed by tracking the stores a user visits, the bookshelves she favors in a library, the specific professor's/doctor's office she visits, and the relics she likes in a museum. The exposure of users' location data not only raises the users' privacy concerns but is also against the law in several countries (\emph{e.g.}, GDPR\footnote{\url{https://gdpr-info.eu}}, HIPPA\footnote{\url{https://www.hhs.gov/hipaa}}, and PDPA\footnote{\url{https://www.pdpc.gov.sg}}).

Given the relevance of the problem, several solutions in the literature have been proposed to protect the users' privacy in indoor locations. These approaches vary from the application of anonymization techniques (\emph{e.g.}, K-anonymity)~\cite{sazdarLowcomplexityTrajectoryPrivacy}\cite{9130897}, encryption protocols \cite{li_infocom}\cite{7417168}\cite{8486221} to Differential Privacy \cite{8493532}. 
However, the proposed methodologies either require the presence of a trusted server who collects individuals' original data or suffer from the computations (and communication) costs. To eliminate the need for a trusted party and the computation overhead, \emph{Local Differential Privacy} (LDP) offers a strong privacy guarantee in which the individuals perturb their data locally (on their own device) before sending them to the third party (named aggregator)~\cite{corr/abs-0803-0924}. It has also shown its efficiency and accuracy in statistical analysis over protected data~\cite{corr/abs-2103-16640}.

In this study, we propose a privacy-aware framework that employs LDP to protect the users' location data while the estimation of the crowd population in different zones of an indoor area remains accurate. This is achieved through the inherent property of LDP in frequency estimation. In this framework, a user's exact location is changed locally before leaving her device.
In addition, we investigate the impact of the indoor location dataset and environmental properties, privacy mechanism, and privacy level on the performance of the proposed framework. We also show that the suggested framework can easily be adapted to different environments as it is not dependent on the number of users and transmitters or their geometric features. The main contributions of this work can be summarized as follows:
\begin{itemize}
    \item We propose a novel privacy-aware LDP-based framework that preserves the location data privacy of people with high adaptability. 
    \item We evaluate the performance of the suggested framework using six different LDP-based privacy mechanisms while considering the effect of privacy levels.
    \item We investigate the impact of indoor environment properties, such as the number of users and transmitters, on the performance of our proposed framework.
\end{itemize}

The rest of this paper is organized as follows. Section~\ref{sec:pre} presents the preliminary concepts of Wi-Fi indoor location fingerprinting and LDP. Section~\ref{sec:method} provides an overview of the proposed framework for LDP-based privacy preservation in indoor localization. The experimental setup, description of datasets, evaluation metrics, and experimental results are presented in Section~\ref{sec:experiment}. Section~\ref{sec:rw} compares our study with the related work. Finally, Section~\ref{sec:conclusion} concludes the paper. 

\section{Preliminaries} \label{sec:pre}
This section explains indoor localization (as well as its common techniques) and LDP (along with its frequency-based estimation algorithms) as considered in this work. The notations used throughout this paper are introduced in Table \ref{tab:notations}.

\begin{figure}[t]
	\centerline{\includegraphics[width=0.5\textwidth]{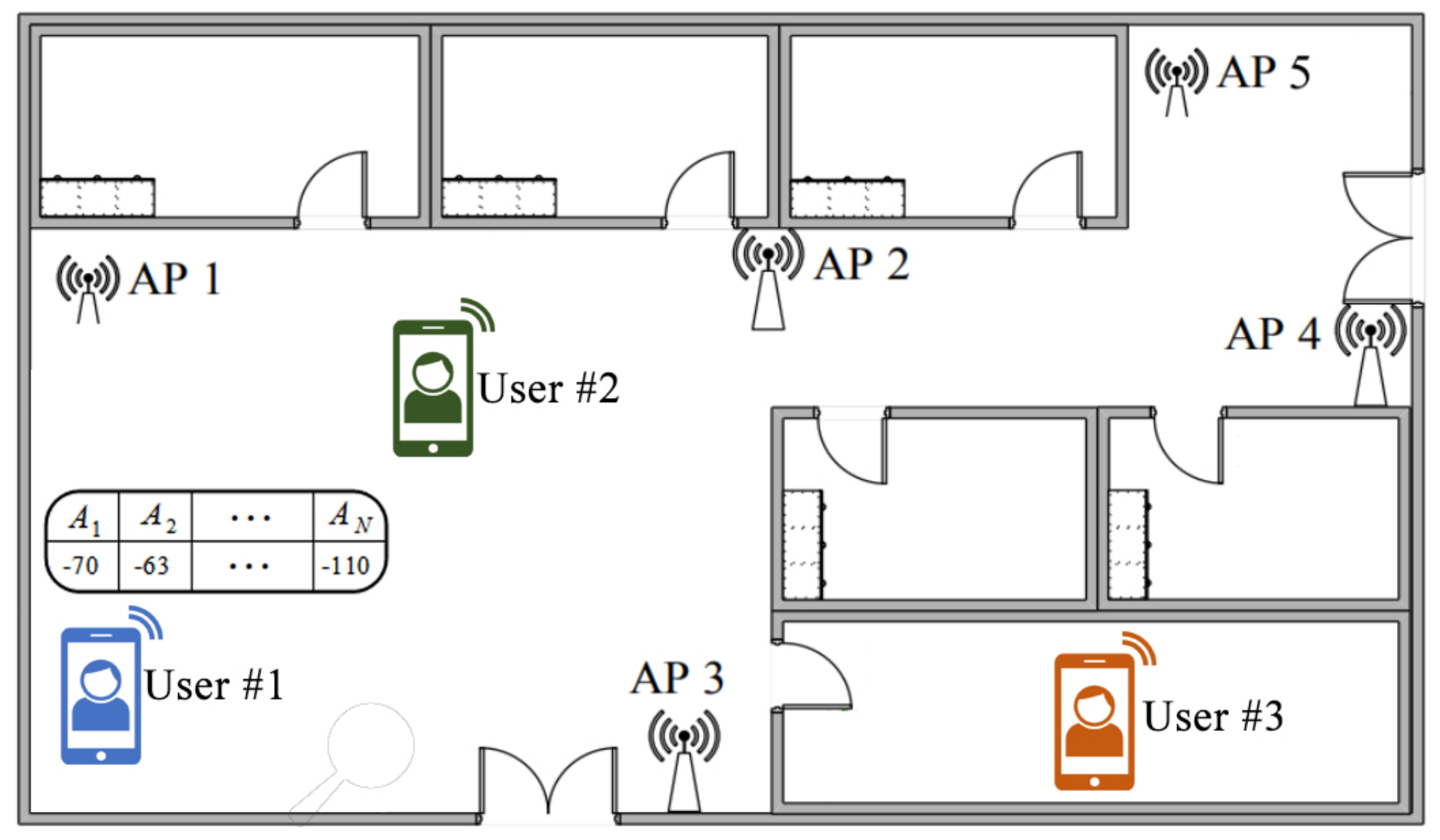}}
	\caption{Overview of indoor location fingerprinting.}
	\label{fig:loc}
\end{figure}

\subsection{Location Fingerprinting} \label{sec:ldp}
Fingerprinting is a cost-effective and high-precision localization technique employed mainly in indoor localization. The basic idea behind this technique is to build a radio map of the indoor environment and employ machine learning techniques to estimate the location~\cite{MoghAccess}. In this method, environmental characteristics are measured and then compared to the existing data. Usually, the fingerprinting method consists of two phases: offline (training) and online (test) phases. In the offline phase, fingerprint data (RSSI) from existing Access Points (APs) or beacons are collected at predefined locations, called Reference Points (RPs). In the online phase, users send their RSSI information from sensed APs to the server. The server can then estimate the user's location using the information collected in both offline and online phases \cite{singhMachineLearningBased}. The schematic of an indoor location fingerprinting is illustrated in Fig. \ref{fig:loc}.

By utilizing fingerprinting, the features of each RP can be mapped to a geographical location. To obtain high positioning accuracy, both temporal and spatial patterns should be considered. One of the main advantages of fingerprinting over other localization methods is that it does not require knowing physical parameters such as distance or angle. Various types of wireless signals can also be used for localization. However, Wi-Fi is the most commonly used signal in fingerprinting. The fingerprint at each RP can be denoted by:
\begin{equation}
    f = \left[(x,y), {{RSSI}_{1}},{{RSSI}_{2}}, \ldots, {{RSSI}_{N}}  \right]
\end{equation}
where $(x,y)$ is the location coordinates of the RP and ${{RSSI}_{i}}$ $\in\mathbb{R}$ is the RSSI value from $i$-th AP at that RP. 

\subsection{Local Differential Privacy (LDP)} \label{sec:ldp}
In LDP, an aggregator collects information from the users who do not fully trust it but are willing to participate in the aggregator's analysis. To protect the confidentiality of data, each user's value is perturbed locally using a randomized algorithm before being sent to the aggregator. The perturbation might require a pre-process algorithm over input data named encoding. The aggregator then collects the perturbed values and estimates the true statistic. Accordingly, the majority of LDP-based algorithms include the following three phases~\cite{corr/abs-2103-16640}: \emph{(i)} \textit{Encode,} where the user encodes her original value $v$ to $v_t$ using a predefined encoding scheme adopted to the perturbation mechanism, \emph{(ii)} \textit{Perturb,} where the encoded value $v_t$ is perturbed by a randomized algorithm that guarantees LDP, and \emph{(iii)} \textit{Aggregate,} where the aggregator collects the perturbed values from all users and aggregates them to estimate the query result (population frequency in this study).

\begin{table*}[!t]
	\centering
	\resizebox{0.7\textwidth}{!}{
		\begin{tabular}[t]{llll}
			\toprule

			\textbf{Notation} &  \textbf{Description} &  	\textbf{Notation}  &  \textbf{Description}  \\
			\midrule
			$N$ & number of APs &   $v$ & a user's  value \\
			$\mathcal{A}$ & set of APs & $v_t$ &  a user's encoded value\\
			$K$ & number of users & $\hat{v_t}$ & a user's perturbed value  \\
			$\mathcal{U}$ & set of users & $\varepsilon$ & privacy level \\
			$R^i$  &  RSSI vector for $i$-th user  & $\mathcal{D}$ & domain of users' values  \\
			$RSSI_j^i$ & RSSI of $j$-th AP for $i$-th user \,\, & $d_m$ & $m$'th domain element\\
			$t$ & time window & $\mathbf{Z}$ & zone binary vector\\
			$\mathbf{R}$ & RSSI matrix & $\Phi $ & perturbation algorithm \\
            $M$ & number of strongest APs  &  $\Psi $ & aggregator estimation algorithm\\                           
            $L$ &  number of created zones & $\hat{f} $ & privacy mechanism\\

			\bottomrule
	\end{tabular}	}
	\caption{Notations. } 
	\label{tab:notations}
\end{table*}

\begin{definition} [$\varepsilon$-Local Differential Privacy ($\varepsilon$-LDP)~\cite{corr/abs-0803-0924}] 
A randomized mechanism $\mathcal{M}$ satisfies $\varepsilon$-LDP if and only if for any pair of input values $v, v' \in \mathcal{D}$ (where $\mathcal{D}$ is the domain set) and for any possible output $S \subseteq Range(\mathcal{M})$, we have 
\begin{align}
    & Pr [\mathcal{M}(v) \in S ] \, \leq \, e^{\varepsilon} Pr[ \mathcal{M}(v') \in S  ]
\end{align} 
when the value of $\varepsilon$ is known from the context, we omit $\varepsilon$ from $\varepsilon$-LDP and simply write LDP.
\end{definition} 
Compared with the setting of DP, the local setting offers a more robust level of privacy since the aggregator has only access to the perturbed data. In this case, even if the aggregator is malicious, individuals' private data is protected by the guarantee of LDP \cite{203872}. The LDP-based mechanisms are mainly defined based on the randomized response as follows. 
\begin{definition} [Randomized Response~\cite{corr/HolohanLM16}]
Let $v$ be a user's binary value, $v_t$ be its encoded version and $\hat{v_t}$ be its perturbed response. Then, for any value $v$,
 \begin{equation}
Pr[ \hat{v_t} = v ] = 
\begin{cases}
 \frac{e^{\varepsilon} }{ e^{\varepsilon} +1}  & \text{if } \quad    v_t = v, \\
 \frac{ 1 }{ e^{\varepsilon} +1}  & \text{if }\quad  v_t \neq v
\end{cases}
\end{equation} 
The randomized response returns the true value with probability $\frac{e^{\varepsilon} }{ e^{\varepsilon} +1  } $ and the random value with probability $\frac{1 }{ e^{\varepsilon} +1} $.
\end{definition}

\subsubsection{LDP-based Frequency Estimation~\cite{corr/abs-2103-16640}:} \label{ldpfrequency}
In frequency estimation, it is assumed that there are $n$ users, such that each user has one value $v$ from the domain set $\mathcal{D}= \{d_1, d_2, \ldots, d_M\}$ and reports the perturbed version once.  
The data aggregator aims to find the number of users having a value $d_m \in \mathcal{D}$ for all values in the set as the following:
\begin{align}
	& \mathcal{N} (d_m) = \frac{ \sum_{i=1}^{n}  1_{\Phi( v_t )} (d_m) - nq }{p - q} \label{eq:frequency}
\end{align}
where $v_t$ is the encoded version of $v$, $\Phi(v_t)$ is the perturbed output of user $u_i$ by perturbation algorithm $\Phi$, and $1_{\Phi(v_t)} (d_m)$ is an indicator function which is equal to 1 when $\Phi(v_t) = d_m$, and 0 otherwise; $p$ and $q$ are the perturbation probabilities. 

\smallskip 
Let $\Phi$ be the randomized method used by the users for perturbing their original values, and $\Psi$ be the estimation method used by the aggregator, then the pair of algorithms $\langle \Phi, \Psi \rangle$ that enables the aggregator to estimate the frequency of a value in a dataset is called a frequency oracle. 
We use a privacy mechanism instead of a frequency oracle in the rest of this study. The privacy mechanisms used in this study are the following ones~\cite{corr/abs-2103-16640}.

\smallskip

\textbf{Optimal Local Hashing (OLH):}
Let $\mathbb{H}=\{ \mathcal{H}_1, \ldots, \mathcal{H}_m \}$ be a universal hash function family such that each $\mathcal{H} \in \mathbb{H}$ outputs a value in $[g]$ for $g \ge 2$, where $[g]$ is a  dataset of size $g$ (where $g$ is rounded up to a natural number for float values). The user's input value $v$ is encoded as $\langle \mathcal{H}, v_t \rangle$, where $\mathcal{H} \in \mathbb{H}$ is chosen uniformly at random and $v_t= \mathcal{H}(v)$.
At perturbation step:
\begin{equation}
Pr[\hat{v_t} = \mathcal{H}(v)] =  \begin{cases}
\frac{e^{\varepsilon} }{ e^{\varepsilon} +g -1  }      & if \,\, \quad  v_t = \mathcal{H}(v)  \\
\frac{1}{ e^{\varepsilon} + g - 1}    &if \,\, \quad v_t \neq \mathcal{H}(v)
\end{cases} 
\end{equation}
where $g =  e^{\varepsilon} +1$.

\smallskip
\textbf{Optimized Unary Encoding (OUE):}
In Unary Encoding, a value is encoded as a bit vector in which the $v$'th component is 1 and the remaining components are zero. Given probabilities $p$ and $q$, the perturbed output is computed as follows:

\begin{equation}
 Pr[\hat{v_t}[i] = 1  ] = \begin{cases}
p & if \,\,\,\quad  v_t[i] =1 \\
q &if \,\,\,\, \quad v_t[i] =0\\
\end{cases}
\end{equation}
The optimal parameters for $p$ and $q$ in OUE are $p = \frac{1}{2}$ and $q = \frac{1}{e^{\varepsilon} + 1}$ which minimize the error. 

\smallskip
\textbf{Thresholding with Histogram Encoding (THE):} 
In Histogram Encoding (HE), an input value $v$ in the domain $\mathcal{D}$ is encoded using a length $|\mathcal{D}|$ histogram, in which the $v$'th component is $1$ and the remaining components are zero. 
This vector is then perturbed using Laplace distribution such that $\hat{v_t}[i] = v_t[i] + Lap (\beta)$ where the probability of selecting a point $x$ is equal to 
$\frac{1}{2 \beta} e^{- |x|/ \beta }$. In THE, a threshold parameter $\theta$ is given as input and each noisy count above the threshold is interpreted as $1$ and below the threshold as $0$.

\smallskip
\textbf{Hadamard Random response (HR):}
In transformation-based methods as HR, a public random matrix $\Phi \in \mathbb{R}^{d' \times d} = \{ -\frac{1}{ \sqrt{d'} } , \frac{1}{ \sqrt{d'} }  \}^{d' \times d}$ is generated by the aggregator, where $h$ is a parameter estimated by the error bound (error is defined as the maximum distance between the estimation and true frequency). 
In the encoding phase, the user's value $v$ is encoded as $\langle v_r, v_x \rangle$, where $v_r$ is randomly selected from $\{1, \ldots, d'\}$ and $v_x$ is the $v$'s element of the $v_r$'s row of $\Phi$, \emph{i.e.}, $v_x = \Phi[v_r,v]$.
In perturbation phase, $\langle  v_r, v_x \rangle$ is perturbed as $\langle v_r, b \cdot e \cdot d'\cdot v_x \rangle$, where $e = (e^{\varepsilon} +1)/ (e^{\varepsilon} -1) $ and
\begin{equation}
b = 
\begin{cases}
+1   \quad & \text{with probability }\ \,\,  p = \frac{e^{\varepsilon}  }{e^{\varepsilon}+1 }  \\
-1   \quad  & \text{with probability}\  \,\,\,   q= \frac{1}{e^{\varepsilon} +1}
\end{cases}
\end{equation}
After collecting the perturbed reports $ \langle v^i_r, v^i_y \rangle$, the aggregator estimates the frequency of value $d_m$ as
\begin{align}
& \hat{f}(d_m) = \sum_{i} v^i_y \cdot \Phi[v^i_r, d_m]
\end{align}
The transformation matrix $\Phi$ in HR is an orthogonal, symmetric $2^M\times 2^M$ matrix in which $\Phi[x,y] = 2^{ -\frac{M}{2} (-1)^{  \langle  x,y \rangle} }$

\smallskip
\textbf{Randomized Aggregatable Privacy-Preserving Ordinal Response (RAPPOR)}: RAPPOR is a hash-based frequency statistical method that randomly selects a hash function $\mathcal{H}$  from a  hash function family  $\mathbb{H} =\{ \mathcal{H}_1, \ldots, \mathcal{H}_m \}$, where each function outputs an integer in $[k]= \{ 0,1, \ldots, k-1 \}$. 
RAPPOR then encodes the hash value $\mathcal{H}(v)$ as a $k$-bit binary vector, and randomized response is performed on each bit. 
Accordingly, the encoded vector $v_t$ is shaped as follows:
\begin{equation}
  v_t[i]=  \begin{cases}
1  &if \,\,\quad  \mathcal{H}(v) =1 \\
0   & otherwise
\end{cases}
\end{equation}
The encoded vector is then perturbed as:
\begin{equation}
Pr[\hat{v_t}[i] = 1  ] =  \begin{cases}
	1- \frac{1}{2}  f    & if \,\, \quad  v_t[i] =1 \\
	\frac{1}{2}  f & if \,\, \quad v_t[i] =0
\end{cases}
\end{equation}
where $f=2 / (e^{ \frac{\varepsilon}{2}}  + 1) $.

The aggregator employs Lasso regression to improve the estimated frequency value out of collected reports. Given that for reducing the communication costs and statistical variance,  the hash function maps the input value with large domain size into a smaller domain, collision problem rises up. Bloom filter~\cite{rappor} and \emph{Count Mean Sketch} (CMS)\footnote{https://machinelearning.apple.com/research/learning-with-privacy-at-scale} 
are two common ways used by Google (RAPPOR) and Apple (CMS), respectively, to reduce the effect of collision. 

\section{Methodology} \label{sec:method}
It is assumed that an indoor environment consists of $N$ active APs and $K$ users, denoted by the sets $\mathcal{A}=\{A_1, \ldots, A_N\}$ and $\mathcal{U}=\{U_1, \ldots, U_K\}$, respectively. Users are able to constantly measure the RSSI of the sensed Wi-Fi signals. Thus, the RSSI from all APs for the $i$-th user ($1\leq i \leq K$) at the time window $t$ can be denoted by the fingerprint vector $R^i=\{RSSI_{1}^{i}, \ldots, RSSI_{N}^{i}\}$, where $RSSI^i_n$ denotes the RSSI value from the $n$-th AP. Since the user might not be able to receive a signal from one or more APs, due to far distance or serious blockage and obstructions, the RSSI values of those APs are set to $-110$dBm in the fingerprint vector of the user. Consequently, we can form an $K \times N$ matrix of RSSI values for the same time window $t$, where each row represents the fingerprint vector of each user:
\begin{equation}
    \mathbf{R}=\left[ \begin{matrix}
   RSSI_{1}^{1} & RSSI_{2}^{1} & \ldots  & RSSI_{N}^{1}  \\
   RSSI_{1}^{2} & RSSI_{2}^{2} & \ldots  & RSSI_{N}^{2}  \\
   \vdots  & \vdots  & {} & \vdots \\
   RSSI_{1}^{K} & RSSI_{2}^{K} & \ldots  & RSSI_{n}^{K}  \\
\end{matrix} \right]
\end{equation}

In a typical indoor localization scenario, users send their fingerprint vectors to the server. However, this would seriously threaten their privacy, as users' sensitive data is exposed to the leakage risk. To protect the users' confidential location information, we suggest a novel privacy-aware framework as presented in Fig.~\ref{fig:model}. The main difference between our proposed framework and the traditional location fingerprinting data aggregation is that users do not share their fingerprints or geo-location data. 

As it can be observed, the proposed framework consists of three main components:

\begin{figure*}[h]
	\centerline{\includegraphics[width=0.75\textwidth]{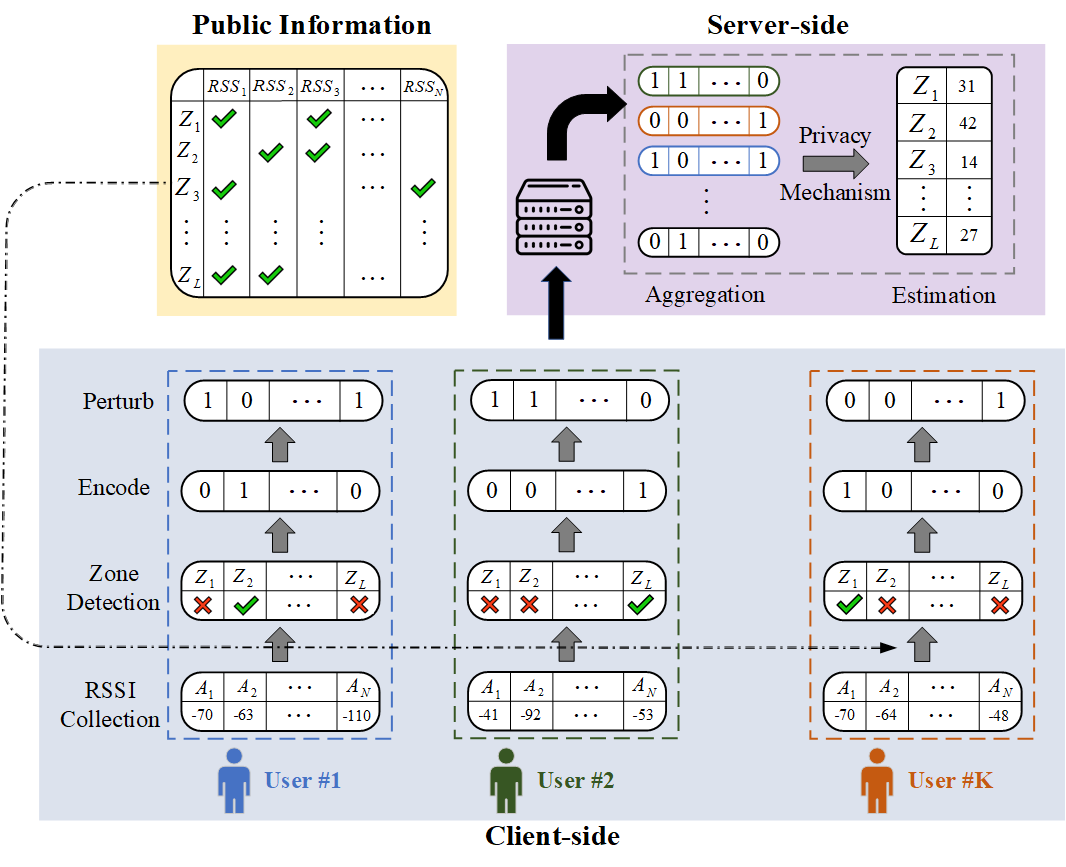}}
	\caption{The block diagram of the proposed system model.}
	\label{fig:model}
\end{figure*}

\smallskip
\textbf{1) Public Information:} In the offline phase, the server utilizes the available fingerprint data to divide the indoor environment into different zones. This zone division method is carried out by measuring the $M$ highest RSSI values in the fingerprint data collected at each RP. In other words, by knowing the $M$ strongest APs at each location, the server can successfully divide the environment into $L$ zones. In theory, the maximum number of possible zones is a function of the total number of APs ($N$) and $M$ strongest APs:
\begin{equation}
{\max(L)={C}_{N,M}}=\frac{N!}{M!(N-M)!}
\end{equation}

In practice, the actual value of $L$ might be less since some combinations might never occur due to the position of APs, the distances between them, and the environmental characteristics of the environment. After the offline phase, the server can safely share the zones and their $M$ corresponding APs as public information with the users. One advantage of this zone division method is that this public information does not reveal the floorplan or the location of APs to the users.

\smallskip
\textbf{2) Client-side:} During the online phase, users first measure the RSSI values from all available APs and then find the $M$ strongest ones. Using the public information shared by the server, hence, users can privately determine their corresponding unique zone. Accordingly, each user forms a binary vector $\mathbf{Z}$ with the size $L$, where:
\begin{equation} \label{eq:zone}
  \mathbf{Z}[j]= \begin{cases}
1  &  \text{if user locates in zone $j$} \\
0  & \text{otherwise}
\end{cases}
\end{equation}

Based on the definition of LDP mechanisms, the user first encodes this binary vector to match the format required by the perturbation step. Next, the user perturbs the encoded data using a perturbation mechanism embedded in the selected privacy mechanism. The perturbed data, which does not contain sensitive information, is then sent to the server. It should be noted that the encoding and perturbation mechanisms might vary from one privacy mechanism to the other one  (Section \ref{sec:ldp}). When utilizing this scenario, the exact zone of each user is hidden to the server, and even the curious server cannot realize the related zone (and consequently the location) of a specific person.
\begin{algorithm}[!t]
  \small
  \KwData{Privacy budget $\varepsilon$,  Privacy mechanism $\hat{f}$,   Public information $\mathbf{T}$.}
   \KwResult{Population frequency of each zone.}
   \For{ \textbf{user} $U_i \in \mathcal{U}$}{
    Measure the $R^i$ from all $N$ APs to shape $R^i=\{RSSI_{1}^{i}, \ldots, RSSI_{N}^{i}\}$. \\
     Determine the user's zone from public $T$ and $R^i$.\\
     Form a binary vector  $\mathbf{Z}^i$ in which $\mathbf{Z}^i[j]$ is 1, if $U_i$ is in zone $j$, and 0 otherwise (Eq.~\ref{eq:zone}). \\
     Encode the binary vector $\mathbf{Z}^i$ to generate the encoded vector $\hat{\mathbf{Z}^i}$.\\
     Perturb the encoded vector $\hat{\mathbf{Z}^i}$  using ${\Phi}_{\hat{f}}$. \\
     Transmit the perturbed vector ${\Phi}_{\hat{f}} (\hat{\mathbf{Z}^i})$ to the \textbf{server}.
   }
   \textbf{Server}: Collects ${\Phi}_{\hat{f}} (\hat{\mathbf{Z}^i})$s and aggregates them using $\Psi_{\hat{f}}$ to estimate the frequency in each zone. \\ 
    \caption{Pseudocode of the proposed framework.} 
  \label{alg:ldpindoor}
\end{algorithm}

\smallskip
\textbf{3) Server-side:} The server aggregates the perturbed data from all users and estimates the population frequency of users in each zone at the time window, $t$. Therefore, the server is able to measure the population in each zone without asking users to share their RSSI fingerprint vectors, which threatens their location privacy. The whole procedure of the proposed framework is presented in Algorithm~\ref{alg:ldpindoor}.

\smallskip
For the privacy analysis of the proposed framework, it is worth mentioning that in this setting, for small $\epsilon$-values even the approximate location of users (\emph{i.e.}, the zone that a user locates in) is protected from the server via perturbation. Moreover, since users do not share any fingerprint data with any third party, their locations cannot be determined through any localization methods. Finally, the proposed framework is invulnerable against adversarial attacks, as the perturbed data of the users being sent to the server does not contain any sensitive information.

\section{Experimental Analysis} \label{sec:experiment}
In this section, we will describe the experiments carried out to evaluate the proposed framework. First, a brief description of the datasets we have used is provided. Next, the experimental setup is introduced as well as the evaluation metrics we have used throughout the experiments. Finally, the experimental results are given.

\subsection{Datasets}

\textbf{CRI dataset:}
This dataset includes the RSSI fingerprints collected in Cyberspace Research Institute (CRI) at Shahid Beheshti University, with an approximate area of 850 $m^2$ \cite{sazdarLowcomplexityTrajectoryPrivacy}. There are nine APs in this dataset, and RSSI fingerprints of 354 users have been captured. 

\noindent
\textbf{JUIndoorLoc dataset:} 
This public dataset contains RSSI fingerprints collected in a multi-floor university building~\cite{royJUIndoorLocUbiquitousFramework}. We use the data of a single floor, with an approximate area of $882$$m^2$, consisting of $36$ APs and $1588$ users. 

Utilizing these two datasets, we can analyze the effect of the dataset size based on the number of both records (users) and APs. General information on the datasets used in our experiment is reported in Table \ref{tab:data}.

\begin{table}[b]
	\footnotesize
	\centering
	\resizebox{0.5\textwidth}{!}{
		\begin{tabular}{lccc}
			\toprule
			\textbf{Dataset} & \, \textbf{Area} \,\, & \,\,  \textbf{$\#$Users} \,\, & \,\, \textbf{$\#$APs}\\
			\midrule
			CRI  & 850 $m^2$  & 354   & 9  \\
			JUIndoorLoc & 882 $m^2$  & 1588  & 36 \\
			\bottomrule
		\end{tabular}
	}
	\caption{Datasets information.}
	\label{tab:data}
\end{table}

\subsection{Experimental Setup}
\begin{figure*}[!t]
	\centerline{\includegraphics[width=0.9\textwidth]{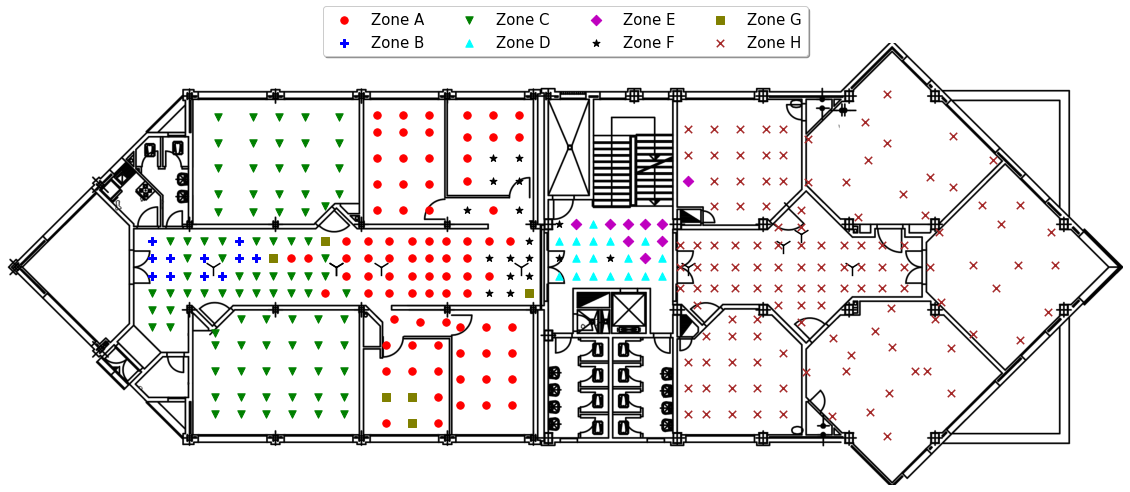}}
	\caption{Eight zones ($L=8$) in CRI environment.}
	\label{fig:zones}
\end{figure*}

We have implemented the privacy-preserving process described in Section \ref{sec:method} using six privacy mechanism, OLH, OUE, THE, HR, CMS, and RAPPOR, in Python. The experiments were performed on a machine running Ubuntu 20.04 LTS with a 64-bit Ryzen 2600x 3.6GHz x6 and 16GB of RAM. The privacy levels ($\varepsilon$) used in $\varepsilon$-LDP are taken from the set $\mathcal{E} = \{0.5, 0.75, 1, 1.5, 2, 3, 5\}$. The input parameters of CMS, RAPPOR, and THE mechanisms have been set to $k_{CMS} =128$, $m_{CMS}=1024$, $k_{RAPPOR} = 64$, $m_{RAPPPOR} = 1024$, and ${\theta}_{THE} = 1$ as proposed in \cite{corr/abs-2103-16640} and \cite{203872}. Inspired by the triangulation methods \cite{singhMachineLearningBased}, we use three strongest APs ($M=3$) for both datasets, which leads the server to divide the environment in the training phase into $L=8$ and $L=26$ zones for CRI and JUIndoorLoc datasets, respectively. However, as mentioned before, any arbitrary value can be chosen for $M$, as long as $M\le N$. The corresponding zone of each user based on the information broadcasting by the server is depicted in Fig.~\ref{fig:zones} for CRI environment.

\subsection{Evaluation Metrics}
 The performance of our proposed framework is evaluated using two distance metrics, namely Root Mean Square Error (RMSE) and Kendall's Tau distance.
 
\begin{definition}
\textit{Root Mean Square Error (RMSE)}
measures the difference between the actual and the estimated number of users in each zone. Formally, the RMSE of two vectors, $z=({{z}_{1}},{{z}_{2}},\ldots ,{{z}_{L}})$ and ${z}'=({{{z}'}_{1}},{{{z}'}_{2}},\ldots ,{{{z}'}_{L}})$ is computed as:  
\begin{equation}
    	RMSE(z,{z}')=\sqrt{\frac{1}{{L}}\sum\limits_{i=1}^{{L}}{{{({{z}_{i}}-{{{{z}'}}_{i}})}^{2}}}}
\end{equation}
Furthermore, RMSE can be normalized as follows:
\begin{equation}
    	NRMSE(z,{z}')=\frac{RMSE(z,{z}')}{{z}_{\max}-{z}_{\min}}
\end{equation}

\end{definition}

\begin{definition}
\textit{Kendall's Tau distance} measures the ordinal association between two ranking lists. For two ranking lists $\tau_1$ and $\tau_2$, the Kendall's Tau distance is defined as:
\begin{align*}
&\kappa(\tau_1, \tau_2) = | \{ (i,j): i <j, \\
& \big(\tau_1(i) < \tau_1(j) \wedge  \tau_2(i) > \tau_2(j)  \big) \\
& \vee  \big(\tau_1(i) > \tau_1(j) \wedge  \tau_2(i) < \tau_2(j)  \big) \}|
\end{align*}
where $\tau_1 (i)$ and $\tau_2 (i)$ are ranks of $i$-th element in $\tau_1$ and $\tau_2$, respectively. The result of $\kappa(\tau_1, \tau_2)$ is zero if the two ranking lists are identical, and $\frac{1}{2} (L(L-1))$ if one list is the reverse of the other one (L is the list size). 
\end{definition}
While RMSE is a good metric for calculating the difference between the real and the estimated population frequencies, Kendall's Tau distance indicates how the order of estimated frequencies differs in the real and the estimated situations. This is beneficial in cases where zones-ordering in terms of population matters (\emph{e.g.}, when the high-populated or low-populated zones are of interest).

\subsection{Experimental Results}
\begin{figure*}[t]
	\begin{subfigure}[b]{0.5\textwidth}
		\rightline{\includegraphics[width=0.9\textwidth]{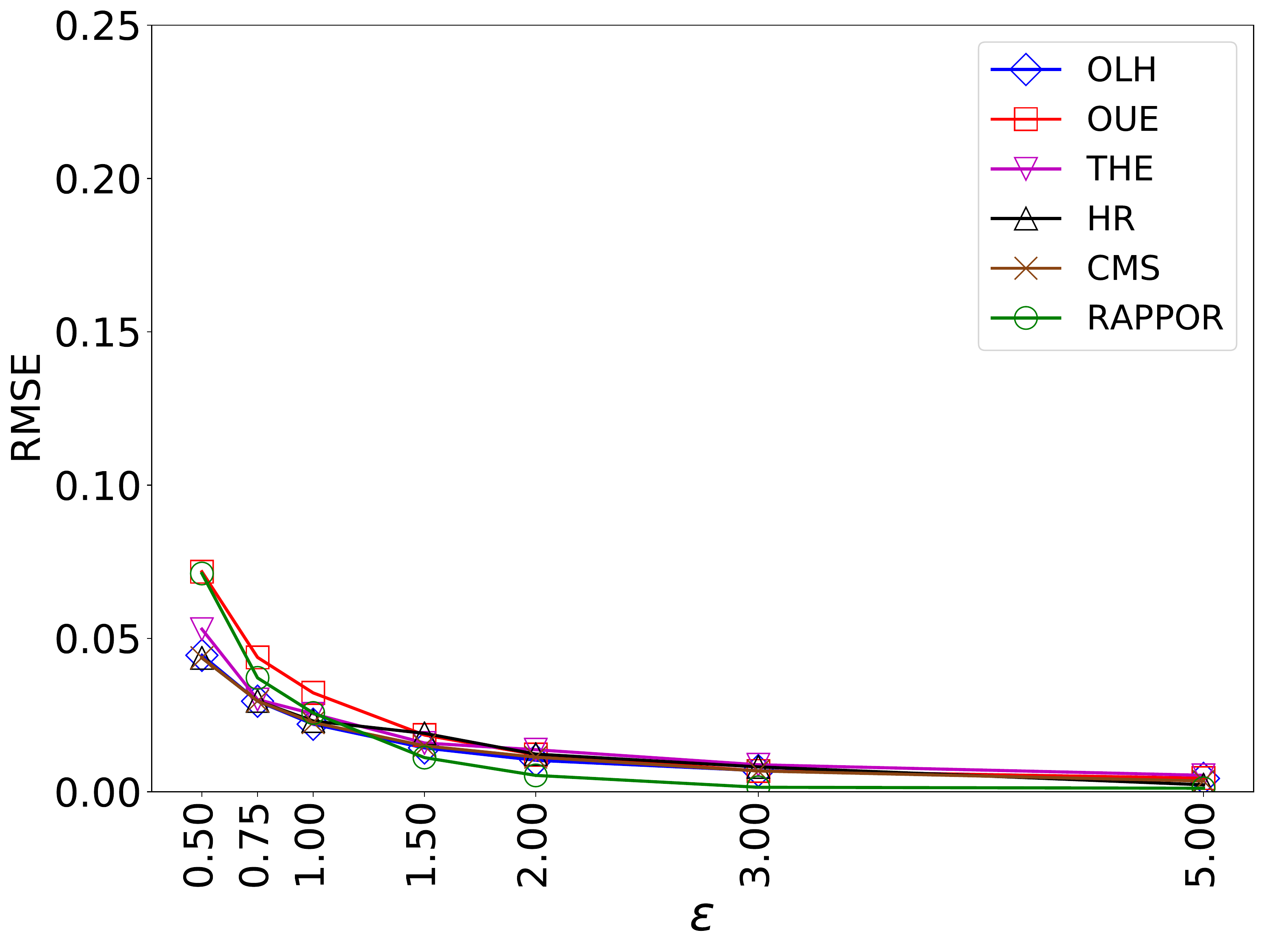} }
		\vspace{-0.25cm}
		\caption{CRI Dataset}	\label{fig:CRI}
	\end{subfigure}
	\begin{subfigure}[b]{0.5\textwidth}
		\rightline{\includegraphics[width=0.9\textwidth]{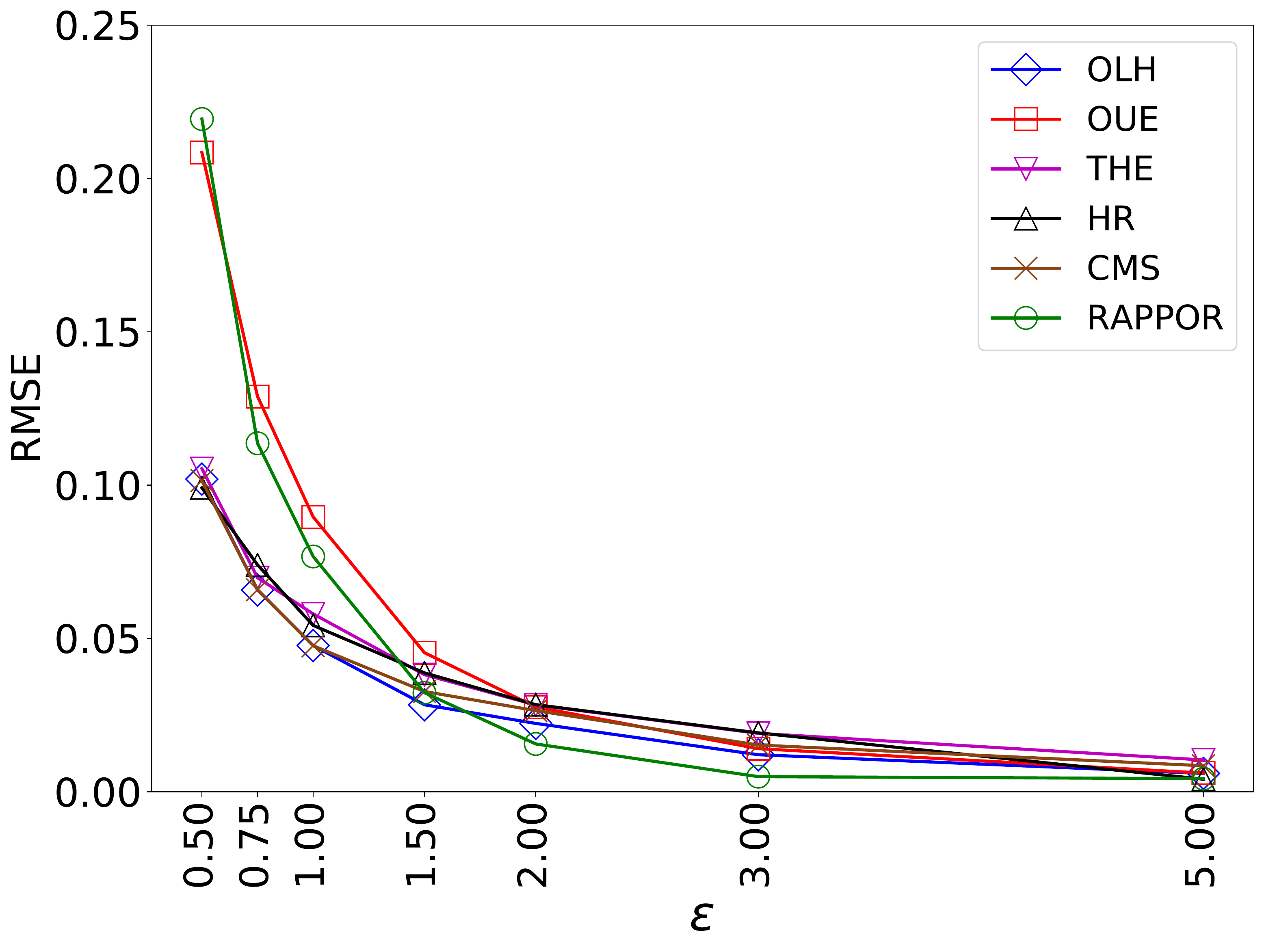} }
		\vspace{-0.25cm}
		\caption{JUIndoorLoc}
		\label{fig:JUIndoorLoc}
	\end{subfigure}
	\vspace{-0.5cm}
	\caption{RMSE of users located in each zone before and after applying LDP mechanism. }
	\label{fig:rmse}
\end{figure*}

\begin{figure*}[!htb]
	\begin{subfigure}[b]{0.5\textwidth}
		\rightline{\includegraphics[width=0.9\textwidth]{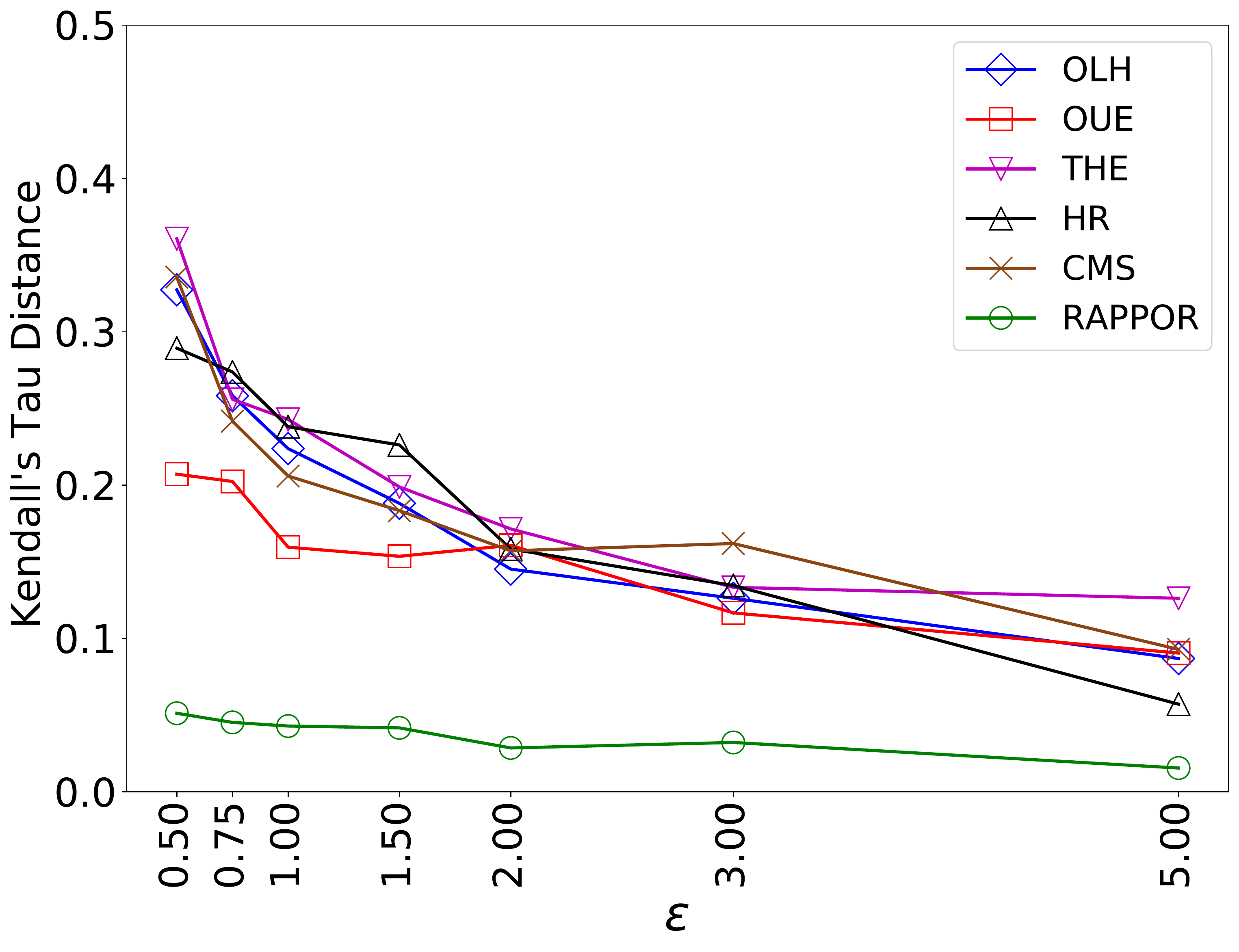} }
		\vspace{-0.25cm}
		\caption{CRI Dataset}	
		\label{fig:CRI_kendall}
	\end{subfigure}
	\begin{subfigure}[b]{0.5\textwidth}
		\rightline{\includegraphics[width=0.9\textwidth]{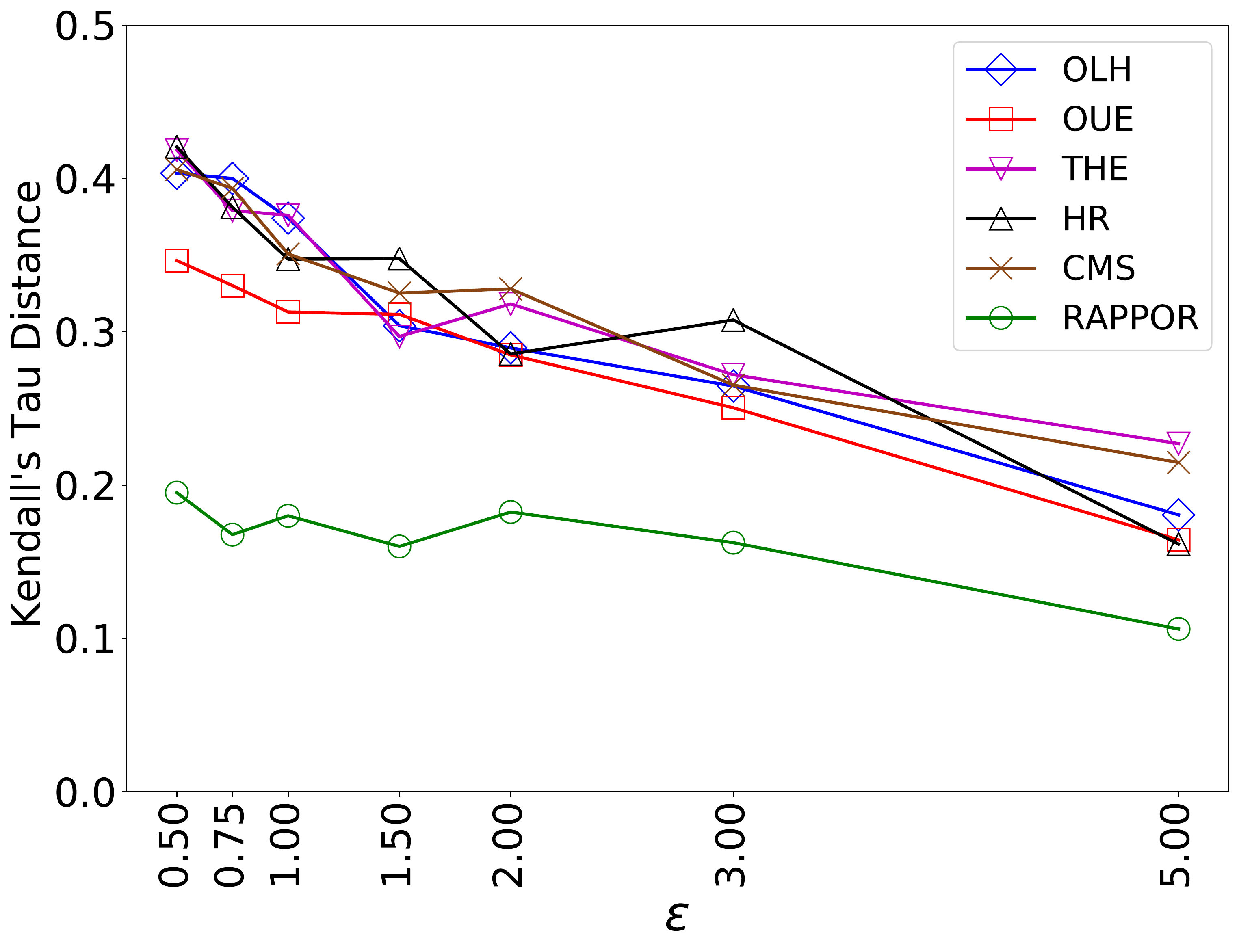} }
		\vspace{-0.25cm}
		\caption{JUIndoorLoc}
		\label{fig:JUIndoorLoc_kendall}
	\end{subfigure}
	\vspace{-0.5cm}
	\caption{Kendall's Tau distance of users located in each zone before and after LDP mechanism.}
	\label{fig:kendall}
\end{figure*}

In order to assess the performance of the proposed privacy-aware framework, we run several experiments, aiming to investigate the impact of three different factors, 1) the dataset properties, 2) the privacy mechanisms, and 3) the privacy level. To this end, we measure the (normalized) RMSE and Kedanll's Tau distances between the actual and the estimated number of users in each zone using six well-known privacy mechanisms with various privacy levels. 
The results are depicted in Fig.~\ref{fig:rmse} and Fig.~\ref{fig:kendall} for both datasets. 
\subsubsection{Impact of dataset:}
In both Fig.~\ref{fig:rmse} and Fig.~\ref{fig:kendall}, a performance gap between two datasets is noticeable. This is mainly because of the difference between the number of zones. As explained in Section \ref{sec:method}, the number of APs and their relative positions and the environmental characteristics (\emph{i.e.}, the placement of walls and other objects) affect the number of zones. However, since LDP-based techniques cannot reduce the effect of random noise over the aggregation of a few inputs, zones with a lower number of users might cause lower performance in frequency estimation. Therefore, the RMSE and Kendall's Tau distance of JUIndoorLoc is slightly higher (worse) than CRI dataset, as the JUIndoorLoc dataset contains zones with only two or three users.

In other words, the size of the dataset does not directly influence the performance of our framework, as long as enough users are presented in each zone.  

\begin{table*}	[!htb]
	\makebox[\textwidth][c]{
		\resizebox{0.95\textwidth}{!}{
			\begin{tabular}{@{ }l rr rr rr rr rr rr rr rr@{ }}
				\toprule
				\multicolumn{1}{l}{\shortstack{\textbf{Privacy} \\ \textbf{Mechanism}}} &
                \multicolumn{2}{c}{\shortstack{\textbf{Zone G} \\ 6}} &
                \multicolumn{2}{c}{\shortstack{\textbf{Zone E} \\ 9}} &
                \multicolumn{2}{c}{\shortstack{\textbf{Zone B} \\ 11}} &
                \multicolumn{2}{c}{\shortstack{\textbf{Zone D} \\ 17}} &
				\multicolumn{2}{c}{\shortstack{\textbf{Zone F} \\ 17}} &
				\multicolumn{2}{c}{\shortstack{\textbf{Zone A} \\ 81}} &
				\multicolumn{2}{c}{\shortstack{\textbf{Zone C} \\ 88}} &
				\multicolumn{2}{c}{\shortstack{\textbf{Zone H} \\ 125}} 
				\\
				\cmidrule(ll){2-3}\cmidrule(ll){4-5}\cmidrule(ll){6-7}\cmidrule(rl){8-9}\cmidrule(ll){10-11} \cmidrule(ll){12-13} \cmidrule(ll){14-15} \cmidrule(l){16-17}
				
				& After & Diff.  &  After & Diff. &  After & Diff.  &  After & Diff.  &  After & Diff.  &  After & Diff.  &  After & Diff. &  After & Diff. \\
				\midrule
				\midrule
				\textbf{OLH}    & 14 & 8  & 11 & 2 & 20 & 9  & 23 & 6 & 17 & 0 & 77 & -4 &  95  & 7  & 121 & -4 \\ 
				\textbf{OUE}    & 17 & 11 & 14 & 5 & 17 & 6  & 23 & 6 & 21 & 4 & 84 & 3  &  100 & 12 & 140 & 15 \\ 
				\textbf{THE}    & 14 & 8  & 14 & 5 & 22 & 11 & 20 & 3 & 23 & 6 & 95 & 14 &  90  & 2  & 136 & 11 \\
				\textbf{HR}     & 17 & 11 & 18 & 9 & 19 & 8  & 24 & 7 & 24 & 7 & 82 & 1  &  87  & -1 & 134 & 9  \\ 
				\textbf{CMS}    & 16 & 10 & 17 & 8 & 18 & 7  & 17 & 0 & 18 & 1 & 86 & 5  &  86  & -2 & 118 & -7 \\ 
				\textbf{RAPPOR} & 10 & 4  & 13 & 4 & 15 & 4  & 22 & 5 & 21 & 4 & 91 & 10 &  99  & 11 & 140 & 15 \\ 
				 $\lceil { \sum \frac{| \textbf{diff} |}{ \textbf{original}} }\rceil  $ &  & 9&  &4 &  & 5&  &2 &  &2 &  & 1&  & 1& &1 \\
				\midrule			 
				\bottomrule
			\end{tabular}
		}
		}
	\caption{Population frequency of each zone in the CRI dataset before and after applying privacy mechanisms for $\epsilon =2$. The numbers under the zones' names are the original population frequencies before applying privacy mechanisms.}
	\label{tab:zones}
\end{table*}
\begin{table*}	[!htb]
	\makebox[\textwidth][c]{
		\resizebox{0.95\textwidth}{!}{
			\begin{tabular}{@{ }l rr rr rr rr rr rr rr rr@{ }}
				\toprule
				\multicolumn{1}{l}{\shortstack{\textbf{Privacy} \\ \textbf{Mechanism}}} &
				\multicolumn{2}{c}{\shortstack{\textbf{Zone A$'$} \\ 2}} &
				\multicolumn{2}{c}{\shortstack{\textbf{Zone B$'$} \\ 3}} &
                \multicolumn{2}{c}{\shortstack{\textbf{Zone F$'$} \\ 6}} &
                \multicolumn{2}{c}{\shortstack{\textbf{Zone E$'$} \\ 7}} &
                \multicolumn{2}{c}{\shortstack{\textbf{Zone D$'$} \\ 48}} &
				\multicolumn{2}{c}{\shortstack{\textbf{Zone C$'$} \\ 67}} &
				\multicolumn{2}{c}{\shortstack{\textbf{Zone G$'$} \\ 199}} &
				\multicolumn{2}{c}{\shortstack{\textbf{Zone H$'$} \\ 976}} 
				\\
				\cmidrule(ll){2-3}\cmidrule(ll){4-5}\cmidrule(ll){6-7}\cmidrule(rl){8-9}\cmidrule(ll){10-11} \cmidrule(ll){12-13} \cmidrule(ll){14-15} \cmidrule(l){16-17}
				
				& After & Diff.  &  After & Diff. &  After & Diff.  &  After & Diff.  &  After & Diff.  &  After & Diff.  &  After & Diff. &  After & Diff. \\
				\midrule
				\midrule
				\textbf{OLH}    & 34 & 32 & 34 & 31 &  27 & 21 & 31 & 24 & 62 & 14 & 80 & 13 & 196 & -3 & 966  & -10    \\ 
				\textbf{OUE}    & 26 & 24 & 32 & 29 &  33 & 27 & 24 & 17 & 53 & 5  & 66 & -1 & 224 & 25 & 1124 & 148   \\ 
				\textbf{THE}    & 42 & 40 & 41 & 38 &  31 & 25 & 33 & 26 & 62 & 14 & 72 & 5  & 199 & 0  & 977  & 1     \\
				\textbf{HR}     & 8  & 6  & 2  & -1 &  9  & 3  & 15 & 8  & 58 & 10 & 80 & 13 & 210 & 11 & 982  & 6     \\ 
				\textbf{CMS}    & 31 & 29 & 33 & 30 &  33 & 27 & 33 & 26 & 54 & 6  & 72 & 5  & 200 & 1  & 990  & 14    \\ 
				\textbf{RAPPOR} & 13 & 11 & 13 & 10 &  16 & 10 & 19 & 12 & 64 & 16 & 84 & 17 & 228 & 29 & 1077 & 101   \\ 
		  $\lceil { \sum \frac{| \textbf{diff} |}{ \textbf{original}} }\rceil $ & & 71 &  & 47 &  & 19&  &17 &  &2 &  &1 &  &1 & & 1 \\
				\midrule			 
				\bottomrule
			\end{tabular}
		}
	}
		\caption{Population frequency of each zone in the JUIndoorLoc dataset before and after applying privacy mechanisms for $\epsilon =2$. The numbers under the zones' name are the original population frequencies before applying privacy mechanisms.}
	\label{tab:zonesjui}
\end{table*}

In order to further study the effect of data size, the population frequencies of zones for both datasets are presented in Tables \ref{tab:zones} and \ref{tab:zonesjui}. Since the number of zones in the JUIndoorLoc dataset is higher ($L=26$), we select eight zones among them: two zones with the minimum user presence (zones A$'$, B$'$), two zones with the maximum user presence (zones G$'$, H$'$), two zones near the mean (zones C$'$, D$'$), and two zones near the median (zones E$'$, F$'$). Furthermore, we chose $\varepsilon = 2$ for all privacy mechanisms as it provides the best balance between privacy and accuracy (knee point in Fig.~\ref{fig:rmse}). These two Tables are sorted from the lower to higher original populations (\textit{i.e.} before applying any privacy mechanism) in zones, where the number of people in each zone has been shown under the name of zones. The difference in population frequency of each zone before and after privatizing are shown for all algorithms. 
The last row of tables reports the average error rate in estimating the number of people in each zone. It can be inferred that the general trend for the average differences over all privacy mechanisms is decreasing when the population in zones is getting higher. Moreover, the overall error for JUIndoorLoc dataset is significantly more than CRI dataset, as it has more zones with very low populations.

\subsubsection{Impact of privacy mechanism:}
According to Fig. \ref{fig:rmse} and \ref{fig:kendall}, RAPPOR mostly outperforms the other privacy mechanisms, especially in ordering the zones based on the population. However, the performance slightly drops for small privacy levels ($\varepsilon \leq 1.5$). In terms of RMSE, the performance of OLH, THE, HR, and CMS is almost identical, especially for larger values of $\varepsilon$. 

\begin{figure*}[!htb]
	\begin{subfigure}[t]{0.48\textwidth}
	\rightline {\includegraphics[width=1\textwidth]{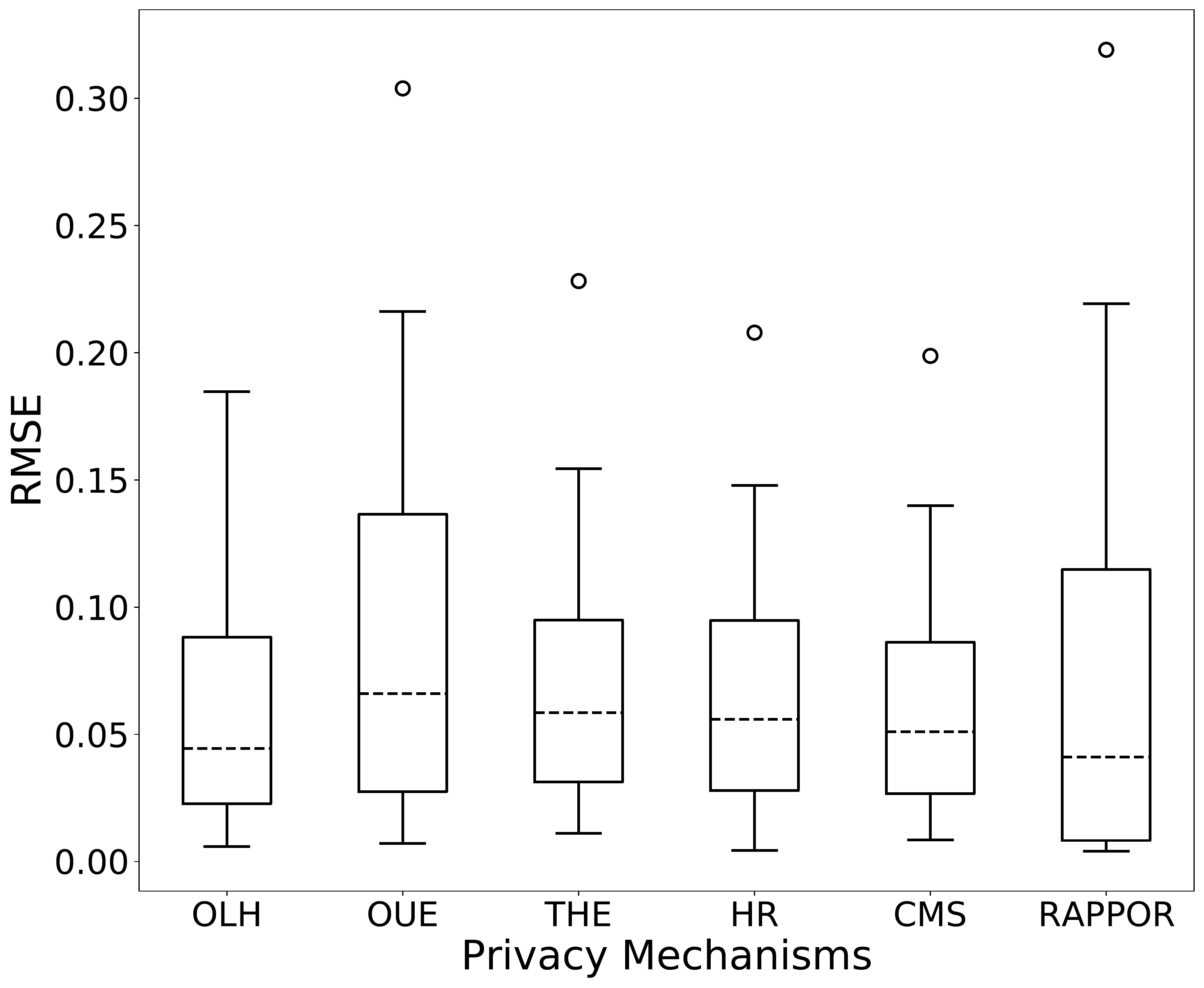} }
		\caption{RMSE distribution.}	\label{fig:rmsebox}
	\end{subfigure}
	\begin{subfigure}[t]{0.48\textwidth}
		\rightline
		{\includegraphics[width=1\textwidth]{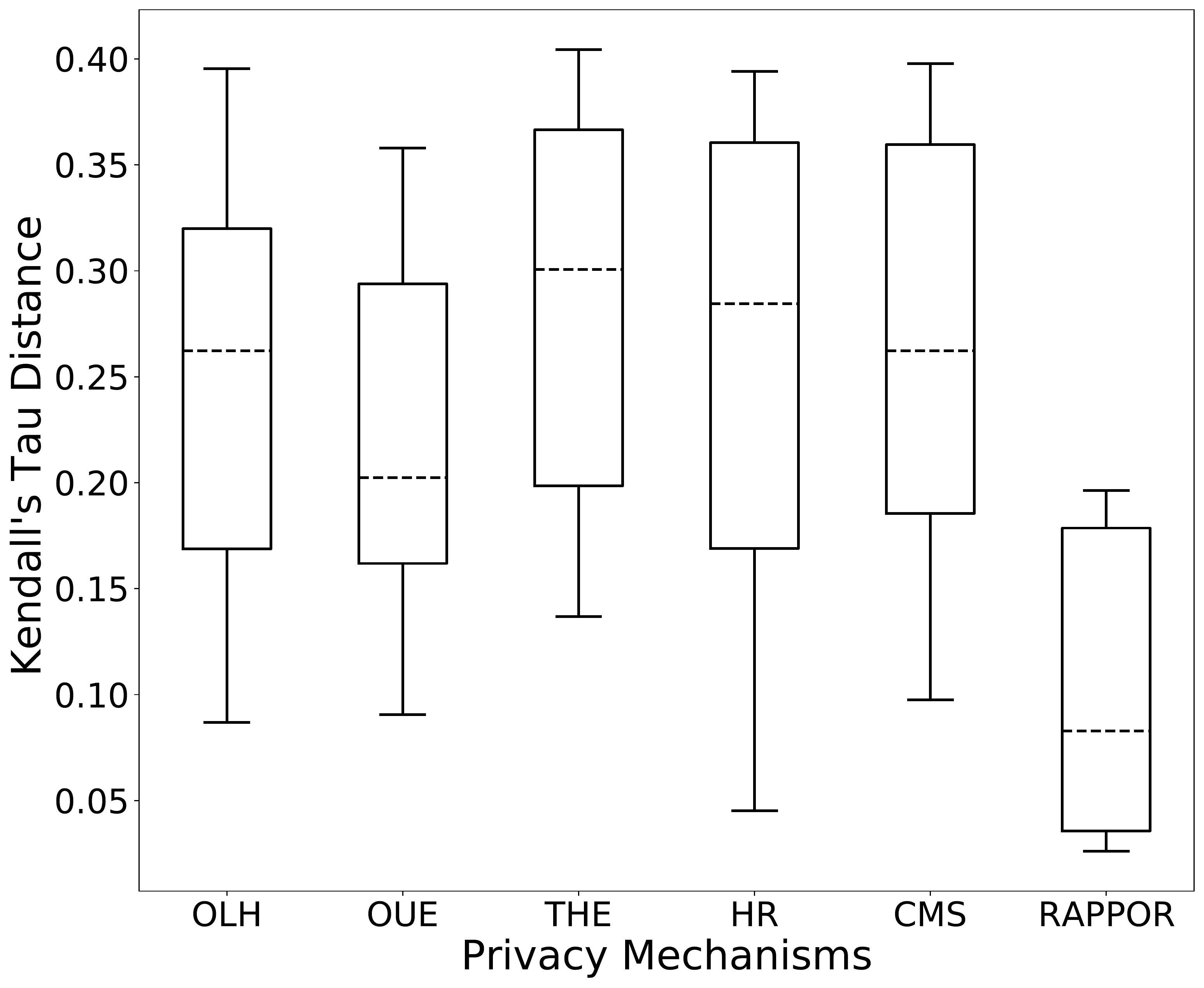}}
		\caption{Kendall's Tau distances distribution.}
		\label{fig:kendallbox}
	\end{subfigure}
		\vspace{-0.2cm}
	\caption{The distribution of aggregated metrics for different algorithms.}
	\label{fig:boxplot}
\end{figure*}

\begin{figure*}[!htb]
			\centering
	\begin{subfigure}[!htb]{0.8\textwidth}
		\centering
		{\includegraphics[width=0.8\textwidth]{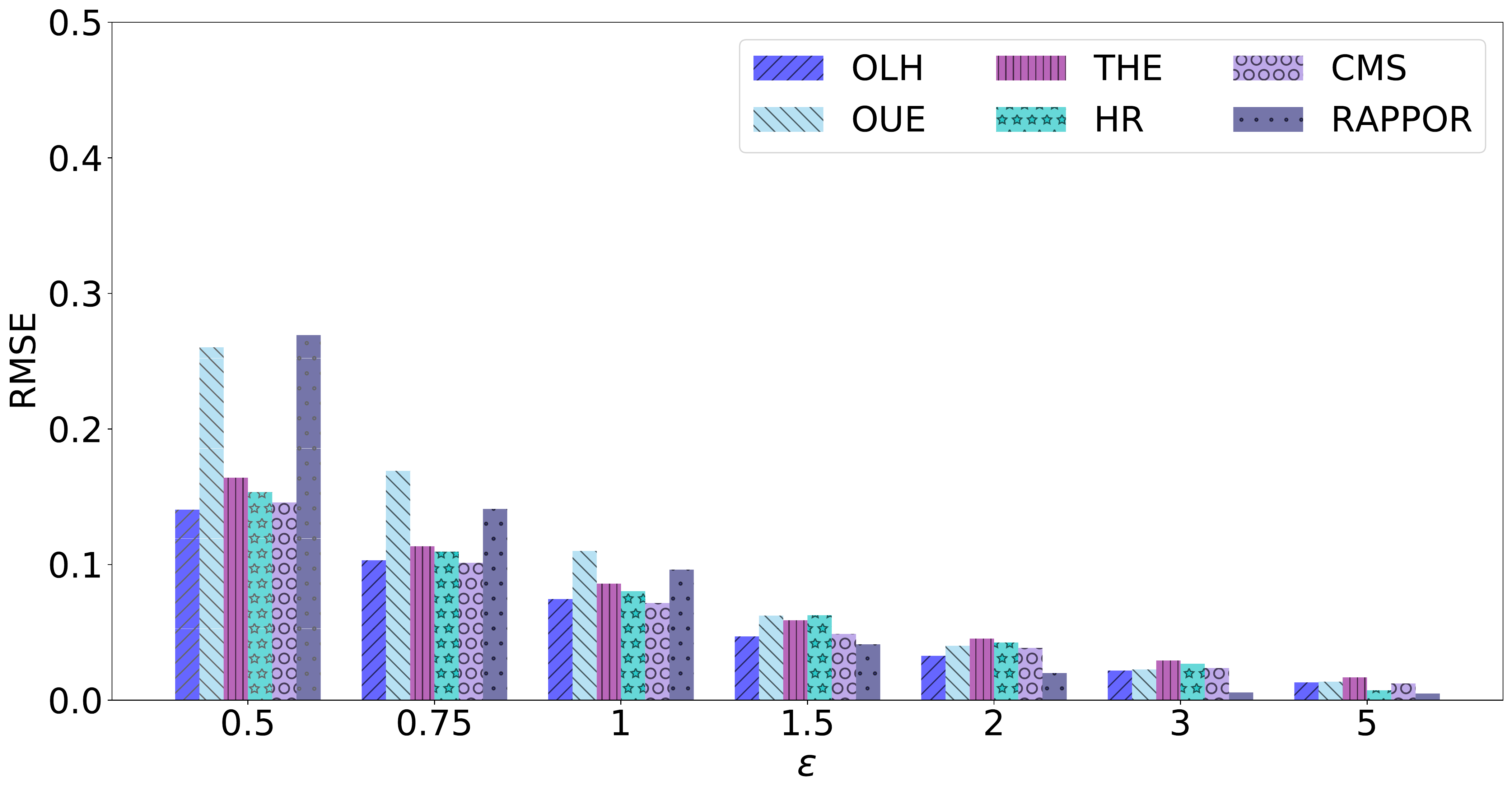} }
		\caption{RMSE}	\label{fig:rmsebarplot}
	\end{subfigure}
\qquad 
	\begin{subfigure}[!htb]{0.8\textwidth}
			\centering
		{\includegraphics[width=0.8\textwidth]{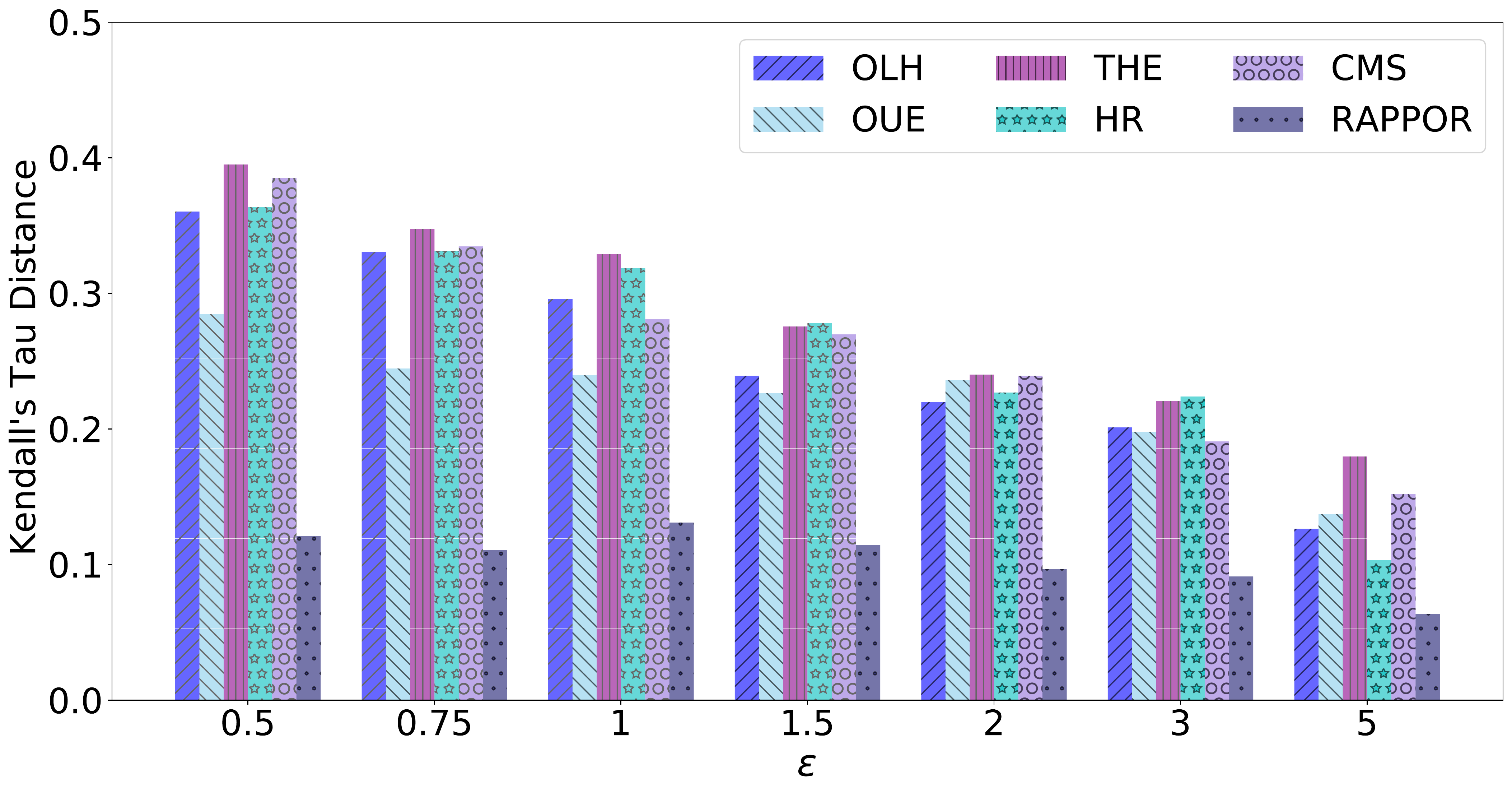} }
		\caption{Kendall's Tau distance}
		\label{fig:kendallbarplot}
	\end{subfigure}
		\vspace{-0.2cm}
	\caption{RMSE and Kendall Tau distance aggregated over both datasets for each privacy level.}
	\label{fig:barplot}
\end{figure*}

We also plot the box-plot of RMSE and Kendall's Tau distance for each privacy mechanism in order to evaluate the performance difference between privacy mechanisms in more detail. To this end, the results of each privacy mechanism are aggregated over both datasets for the mentioned $\varepsilon$ values. The resulting distribution is depicted in Fig.~\ref{fig:boxplot}. Each box represents the distribution of the scores in terms of RMSE and Kendall's Tau distances for the associated privacy mechanism. From Fig.~\ref{fig:rmsebox}, it can be inferred that almost all privacy mechanisms produce comparable RMSE results. Bigger box sizes of RAPPOR and OUE result from these two algorithms' diverse performance for small and large values of $\varepsilon$. The CMS slightly outperforms other algorithms as its box stands at a lower position (with less deviation) compared to other mechanisms. From Fig.~\ref{fig:kendallbox}, it can be observed that in terms of preserving the order of zones, RAPPOR significantly outperforms the other mechanisms as the corresponding box is far lower than the other boxes. 

\subsubsection{Impact of privacy level:}
To gain a better insight on the impact of privacy level, we aggregated the results of RMSE and Kendall's Tau distances over both datasets for each privacy level as depicted in Fig. \ref{fig:barplot}. As expected, as the privacy level increases, \emph{i.e.}, when the privacy constraint is more relaxed, the error has a general decreasing trend. It can also be observed that for small values of privacy level ($\varepsilon \leq 1.5$), even though OUE and RAPPOR are able to keep the population order of zones, they show poor performance similar to the other mechanisms in terms of accuracy (RMSE). For higher privacy values ($\varepsilon \ge 3$), the proposed framework produces optimistic results (comparable to non-private settings) regardless of the privacy mechanism and dataset.

\section{Related Work} \label{sec:rw}
Collecting geospatial data is beneficial for location service providers, as it enables them to increase their quality of service. However, since these data contain sensitive information about private activities, collecting the raw data can potentially leak personal information. Different privacy-preserving approaches have been employed to protect the users' exact location in an indoor environment to address this issue. 

Cryptographic-based techniques by providing a strong privacy guarantee have been used in several studies to encrypt the indoor location data. In \cite{li_infocom}, Homomorphic Encryption is applied to encrypt the RSSI vector of each user. In \cite{7417168}, Fuzzy logic and Homomorphic Encryption are integrated to protect not only the users' data, but also APs information. To alleviate the weaknesses of previous approaches, Paillier public-key encryption is used in \cite{8486221} to protect both users' information and the database stored at the server against attacks. While the encryption-based mechanisms generally provide a strong privacy guarantee, they suffer from severe computation and communication costs~\cite{ZAKHARY2018}. 

Among different anonymization-based approaches, $k$-anonymity is the most commonly used privacy-preserving technique that cloaks a user within the group of $k$ distinct users, such that the user's identity could be well protected against the attacker~\cite{FATHALIZADEH2022102665}. For instance, authors in \cite{sazdarLowcomplexityTrajectoryPrivacy} and \cite{9130897} generate $k-1$ other \emph{dummy} locations to hide a user' exact location among some other fake locations. However, these algorithms are prone to correlation attacks and cannot preserve the semantic information of users' visited locations~\cite{datasharing}. Also, it adds undesirable fake information to real data, which might mislead the system's decision-making process.
 
Differential Privacy (DP), on the other hand, alleviates the shortcomings of previous anonymization techniques by adding systematic noise to the database. For instance, in \cite{8493532}, DP is embedded inside the proposed framework, including four phases: AP fuzzification and location retrieval on the client-side, and DP-based finger clustering and finger permutation on the server-side. The DP-based methodologies, however, suffer from the need for a trusted party in the middle. 
  
LDP has served as an effective solution in the literature in order to eliminate the necessity of a trusted party. For instance, authors in \cite{stephaniePrivacypreservingLocationData} study the case of location-based data clustering by the integration with the edge and cloud computing. Similarly, an LDP-based successive point of interest recommendation system is suggested in \cite{baoSuccessivePointofInterestRecommendation} in which geographical influence, temporal influence, and transition patterns between two points of interest are exploited. Hong et al. \cite{hongCollectingGeospatialData} also investigate the problem of collecting user location data and propose a perturbation mechanism to reduce the error of each collected location data. Finally, authors in \cite{xiongRealtimePrivateSpatiotemporal} investigate the privacy concerns of real-time spatio-temporal data aggregation and suggests using LDP-based algorithms. Publishing and sharing location statistics are also proposed in \cite{erroundaContinuousLocationStatistics} employing LDP to ensure users' privacy while continuously releasing statistics over infinite streams. A probabilistic top-down partitioning algorithm, called LDPart, is also suggested in \cite{zhaoLDPartEffectiveLocationRecord} to generate sanitized location-record data. It utilizes an adaptive user allocation scheme with a series of optimization techniques to improve the accuracy of the released data. 

However, the aforementioned LDP-based studies focus on outdoor location data and do not analyze the inherent property of LDP-based mechanisms in frequency estimations for indoor location data. {They also rely on adding Laplacian (or similar) noise to location-based data. This noise reduces the accuracy of estimating users' population in specific areas.}

For indoor location data, authors in \cite{8736750} and \cite{8253434} propose an LDP-based framework for privacy-preserving indoor location data collection. Our work differs from them in three different parts: we inject location data to the LDP-based framework using the proposed zone division methodology, which does not rely only on one dominant signal; we analyze the performance of several privacy mechanisms in aggregating location data, and our framework is flexible to different environments with a variable number of APs and users.

\section{Conclusion} \label{sec:conclusion}
In this paper, we propose a novel privacy-aware LDP-based framework to collect indoor location data from users while preserving their privacy. Utilizing this framework, users do not share their location-related data since they apply LDP to their data; hence, their locations are kept hidden from the server. Moreover, the presented framework is invulnerable against adversarial attacks, as the perturbed data of the users being sent to the server does not contain any sensitive information. Furthermore, we investigate the influences of the database properties, the privacy mechanism, and the privacy level on the performance of our framework. Experimental results indicate that the presented framework can maintain users' privacy with minimal error. It also provides the population frequency of different zones without knowing the exact location of people inside the indoor area. The proposed method is scalable and easy to implement, making it a solid alternative to other privacy-preserving methods. For future work, we plan to investigate the application of LDP to protect users' privacy over trajectory data.

\bibliographystyle{splncs04}
\bibliography{ref}

\end{document}